\documentclass[twocolumn,twocolappendix]{aastex63}
\newcommand{\tdyn}{t_{\rm dyn}}
\newcommand{\tnot}{t_0}
\newcommand{\ta}{t_{\rm a}}
\newcommand{\Angstrom}{\mathring{\rm A}}
\newcommand{\tautilde}{\tilde{\tau}}
\newcommand{\sinc}{\,\text{sinc}}

\usepackage{amsmath}

\shortauthors{B. Margalit}
\shorttitle{FBOT Light Curves}
\graphicspath{{./}{figures/}}

\begin{document}

\title{Analytic Light Curves of Dense CSM Shock Breakout and Cooling}

\email{benmargalit@berkeley.edu}

\author[0000-0001-8405-2649]{Ben Margalit}
\altaffiliation{NASA Einstein Fellow}
\affiliation{Astronomy Department and Theoretical Astrophysics Center, University of California, Berkeley, Berkeley, CA 94720, USA}



\begin{abstract}
Dense circumstellar material (CSM) is thought to play an important role in observed luminous optical transients: if such CSM is shocked, e.g. by ejecta expelled from the progenitor during core-collapse, then radiation produced by the shock-heated CSM can power bright UV/optical emission. If the initial CSM has an `outer edge' where most of the mass is contained and at which the optical depth is large, then shock breakout---when photons are first able to escape the shocked CSM---occurs near this outer edge. The $\sim$thin shell of shocked CSM subsequently expands, and in the ensuing cooling-envelope phase, radiative and adiabatic losses compete to expend the CSM thermal energy. Here we derive an analytic solution to the bolometric light-curve produced by such shocked CSM. 
For the first time, we provide an analytic solution to the cooling-envelope phase that is applicable starting from shock-breakout and until the expanding CSM becomes optically-thin.
In particular, we account for the planar CSM geometry that is relevant at early times and 
properly treat radiative losses within this planar phase.
We show that these effects can dramatically impact the resulting light-curves, particularly if the CSM optical depth is only marginally larger than $c/v_{\rm sh}$ (where $v_{\rm sh}$ is the shock velocity). This has important implications for interpreting observed fast optical transients, which have previously been modeled using either computationally-expensive numerical simulations or more simplified models that do not properly capture the early light-curve evolution.
\end{abstract}



\section{Introduction} 
\label{sec:intro}

Fast Blue Optical Transients (FBOTs) are an emerging class of optical transients characterized by short rise/fall times ($\sim$days) and large $\gtrsim 10^{43} \, {\rm erg \, s}^{-1}$ luminosities that peak in the blue/UV bands \citep{Drout+14,Rest+18,Pursiainen+18,Margutti+19,Perley+19,Coppejans+20,Ho+20,Ho+20b,Ho+21,Perley+21}.\footnote{This class of events is also often referred to as Fast Evolving Luminous Transients (FELTs). In the present work we adopt the FBOT nomenclature.}
The fast timescales and high luminosities imply that the peak of these events cannot be powered by radioactive decay of $^{56}$Ni, the primary energy source of SNe Ia and many core-collapse SNe \citep{Drout+14}. 
Instead, it has been proposed that the collision of fast-outflowing debris (`ejecta') expelled by the stellar progenitor, e.g. during core-collapse, with dense circumstellar material (CSM; e.g. \citealt{Quataert&Shiode12,Fuller&Ro18}) may power observed FBOT emission.
Within this interaction paradigm, the observed emission is produced as thermal energy generated behind the ejecta-CSM shock diffuses outwards from within the initially opaque CSM.
This scenario may also be relevant in powering superluminous supernovae (SLSNe; \citealt{Quimby+07,GalYam12}), in particular the hydrogen-rich SLSN-II subclass.

To reproduce the rapid decay timescale, a dense CSM with an abrupt outer `edge' has often been invoked \citep{Ofek+10,Chevalier&Irwin11,Ginzburg&Balberg12,Rest+18}.
If the CSM optical depth is sufficiently large ($\tau \gtrsim c/v_{\rm sh}$, where $v_{\rm sh}$ is the shock velocity) near the outer edge $R_0$, then radiation first breaks out from this material when the shock approaches $R_0$. In this case, the optical peak is dominated by cooling-envelope emission of the expanding hot CSM rather than continued shock interaction (the `compact' wind scenario of \citealt{Chevalier&Irwin11}; see also \citealt{Ginzburg&Balberg14,Haynie&Piro21}).
Numerical simulations of this process have shown good agreement with observations \citep[e.g.][]{Ginzburg&Balberg12,Rest+18}, as do analytic estimates of the peak luminosity and duration of transients produced by such configurations. However a full analytic model of the expected light-curves, from shock breakout through the cooling-envelope phase has been lacking.
The aim of this work is to present such a model.

We begin by describing the physical configuration we consider, and discussing qualitative aspects of our model with respect to previous work (\S\ref{sec:description}). In \S\ref{sec:diffusion} we derive the primary result of this work---an analytic model for the time-dependent bolometric luminosity of dense-CSM-powered shock-cooling transients. This derivation follows the classical work of \cite{Arnett80} (see also \citealt{Pinto&Eastman00}), but is tuned to the CSM scenario of interest. In particular, we account for geometric aspects that are relevant in this new scenario (e.g. the planar phase), and show that these can affect the resulting light-curve.
In \S\ref{sec:properties} we discuss properties of the solution and the range of light-curve morphologies that are permitted by it.
In \S\ref{sec:observations} we discuss observational consequences of our model, show approximate band-limited optical light-curves, and compare our results with previous models.
We conclude in \S\ref{sec:discussion} by discussing implications, shortcomings, and directions for future work.
In Appendix~\ref{sec:one-zone} we also derive a simplified `one-zone model' of the light-curve, whose results are compared with the more detailed calculations of \S\ref{sec:diffusion} in \S\ref{sec:properties}.

\section{Model Description}
\label{sec:description}

We consider the following physical setup: a spherical ejecta expanding with velocity $v_0$ runs into a shell of circumstellar material of initial width $\Delta R_{\rm csm}$ located at a radius $\sim R_0$, and whose initial optical depth $\tau_0$ is $> c/v_0$. We consider the CSM to be at rest, but our results apply to any case where the CSM velocity is $\ll v_0$.
For simplicity, we consider the case of a top-hat (constant density) CSM profile that has a sharp outer edge. As discussed below, this reasonably approximates more-complex density profiles so long as most of the CSM mass is contained near the outer edge, and that the optical depth at $\sim R_0$ is $> c/v_0$. 
In particular, this applies also to the case of a truncated wind profile with $\rho = D r^{-2}$ for $r<R_0$ if the wind is `compact', that is if $R_d > R_0$ where $R_d \sim \kappa D v_0 / c$ is the radius at which the wind optical depth would (neglecting the truncation at $r \sim R_0$) equal $\sim c/v_0$ \citep{Chevalier&Irwin11}. In this case, the appropriate ``shell'' mass and width that correspond to the wind scenario are $M \approx 4\pi D R_0$ and $\Delta R_{\rm csm} \approx R_0$.

As the ejecta sweeps the CSM, a shock is formed and part of the ejecta's kinetic energy is dissipated into post-shock CSM thermal energy.
If the optical depth through the CSM is $\gtrsim c/v_0$ then radiation is initially trapped within the flow, and the shock will be mediated by radiation pressure. The flow is then well-described by an adiabatic index $\gamma = 4/3$, and the Rankine-Hugoniot conditions dictate that the shock velocity is $v_{\rm sh} = 7 v_0/6$, and the post-shock density (energy density) is $\rho_0 = 7 \rho_{\rm csm}$ ($e_0 = 7 \rho_{\rm csm} v_0^2 / 2$).
The CSM shell is thus accelerated, heated, and swept by the shock into a thin shell of width $\Delta R_0 \equiv \Delta R_{\rm csm} / 7 \ll R_0$. The thinness of this post-shock region supports our treatment of the CSM as having $\sim$uniform density.
Note that if $\tau_0 < c/v_0$ then a radiation mediated shock cannot form, and the resulting shock emission peaks in the hard X-ray rather than optical/UV band. This situation has been treated in recent work (\citealt{Margalit+22}; see also \citealt{Chevalier&Irwin12,Svirski+12}). Here we focus instead on the case where $\tau_0 \gtrsim c/v_0$.

When the shock reaches a fractional depth $\hat{x} \sim 1 - c/\tau_0 v_{\rm sh}$ from the CSM outer edge (where 
$\tau_0 \approx \kappa \rho_0 \Delta R_0$ 
is the optical depth through the entire CSM and $\kappa$ the opacity), radiation first begins to leak out, producing the so-called shock-breakout emission.
The luminosity of this emission (neglecting light travel time effects) is $\sim E_{\rm bo} / t_{\rm bo} = L_{\rm sh}$, where $E_{\rm bo} \sim 4\pi R_0^2 \Delta R_0 (1-\hat{x}) e_0$ is the thermal energy at $x > \hat{x}$, and $t_{\rm bo} \sim (1-\hat{x})\Delta R_0/v_{\rm sh}$ is the radiative diffusion timescale of the layer. It is equal to the kinetic power of the shock, $L_{\rm sh} = 2\pi R_0^2 \rho_{\rm csm} v_{\rm sh}^3$.

Shock-breakout emission has been studied extensively in the literature (e.g. \citealt{Chevalier92,Nakar&Sari10,Piro+10,Katz+10,Waxman&Katz17}), in particular in cases where the shock propagates down a polytropic density profile $\rho \propto (R_0-r)^n$, as relevant in stellar atmospheres.
In this situation, energy conservation implies that the shock accelerates as it moves outward through regions of decreasing density, and the shock velocity is well-described by the Sakurai self-similar solution, $v_{\rm sh} \propto \rho^{-\mu}$, where $\mu(\gamma,n) \approx 0.19$ \citep{Sakurai60}.
Shock acceleration launches a rarefaction wave that propagates inward (in Lagrangian sense) through the shocked material, and converts an order unity fraction of the region's thermal energy into kinetic energy. The details of this process are complicated, but the resulting flow (after the rarefaction wave crosses the material) is well approximated as homologous ($r \propto v$) where each parcel of shocked gas attains a terminal velocity that is roughly $\sim$two times larger than the velocity of the shock at the same mass coordinate \citep[e.g.][]{Ro&Matzner13}. This yields a self-similar density profile that falls as a steep power-law, $\rho \propto v^{-k}$ with $k \sim 10$ \citep{Chevalier&Soker89,Matzner&McKee99}.

\cite{Nakar&Sari10} studied the emission properties of shock breakout from the fastest ejecta material, that are well-described by the self-similar solution discussed above. A consequence of the steep density profile is that the `luminosity shell' $\hat{x}$ which sets the radius from which breakout emission is observed, propagates into the ejecta as a function of time \citep[see also][]{Chevalier92}. The \cite{Nakar&Sari10} model therefore predicts a slowly declining light-curve since deeper (/denser) layers of material are uncovered as a function of time, and these are able to radiate more energy, compensating (at least partly) for adiabatic losses.

This situation, however, is only applicable within the steep outer ejecta.
\cite{Chevalier&Soker89} show that interior regions of SNe ejecta are instead described by a $\sim$flat density profile ($k \approx 1$) where most of the mass and energy are located at large radii \citep[see also][]{Matzner&McKee99}. Physically, this occurs because the initial stellar density profile is not well-described by $\rho \propto (R_0 - r)^n$ at $r \ll R_0$, and the planar Sakurai solution is no longer applicable.
A flat inner density profile has dramatic effects on the continuous emission. Naive scalings would imply that the luminosity shell propagates {\it outward} once it enters this portion of the density profile \citep{Chevalier92}. However this is impossible, since at larger radii there is a density inflection and $k \gg 1$, in which the luminosity shell propagates inward in Lagrangian sense. This tension is settled by having the luminosity shell fixed to the transition radius at which the density inflection occurs. This point was also recently discussed by \cite{Piro+21}.
Once the luminosity shell remains fixed at the characteristic (density-inflection) ejecta radius, emission is no longer described by the self-similar solutions derived by \cite{Nakar&Sari10}. Instead, the `cooling-envelope' solutions first derived by \cite{Arnett79,Arnett80} are more appropriate. 

In the following section we present and solve the set of equations that describe radiative diffusion and the resulting luminosity in the cooling-envelope scenario. We are motivated to consider this scenario because the bulk of the shocked-CSM density profile should be described by this $\sim$flat portion so long as the initial outer CSM edge is very sharp. In our idealized top-hat CSM model, one can think of the density profile as being described by $\rho \propto (R_0 - r)^n$ with $n \to 0$. In this limit, $\mu \to 1/\left( 2 + 2 \sqrt{2}\right) \simeq 0.207$ and the terminal outer density profile is infinitely steep, $\rho \propto v^{-k}$ with $k = \mu/n \to \infty$ \citep{Ro&Matzner13}. Correspondingly, the mass (and energy) contained within this steep component are infinitesimally small, which motivates us to neglect this component in our analysis.

We define time $t=0$ as the time at which the shock reaches the outer CSM edge. At this time, the CSM is confined to a thin shell of width $\Delta R_0 \equiv \Delta R_{\rm csm} / 7$, whose inner edge is defined to be at $r=R_0$. We assume that the CSM expands homologously at $t>0$, where the velocity of material at the inner edge of the CSM is $v_0$, and the outer velocity is $v>v_0$. This neglects details of the initial acceleration and rarefaction wave that act to establish such a profile, and whose details may influence the flow at times $\lesssim t_0 \equiv \Delta R_0 / v$, the initial expansion timescale. We have also implicitly assumed that the ejecta that runs into the CSM is not decelerated noticeably by this interaction, hence its velocity $v_0$ is fixed, and sets the inner boundary condition for the shocked CSM (which cannot have velocity less than $v_0$). This is akin to the approximation that $M_{\rm csm} \ll M_{\rm ej}$, and therefore the ejecta can be treated as a ``piston'' (this also neglects the ejecta's own density profile, and the reverse shock that would propagate into it, a point we return to in \S\ref{sec:discussion}).
For convenience, we define $\beta \equiv v_0/v$ as the ratio of inner (minimum) to outer (maximum) velocities of material within the shell. In the present case of interest, we expect $\beta \gtrsim 1/(1+2\sqrt{2}) \simeq 0.26$, where this minimum value of $\beta$ is set by the maximum terminal velocity $v$, as determined from the 1D (planar) Riemann rarefaction problem (e.g. \citealt{ZeldovichRaizer} \S I.28). In reality, the maximum velocity will be smaller due to radiative losses, and we adopt $\beta = 0.5$ as a fiducial value.
Finally, we define the dynamical timescale $t_{\rm dyn} \equiv R_0/v$ as the time it takes the CSM to double in radius, and the ``Arnett'' diffusion timescale as $t_a \equiv \left( 3 \kappa M / 4\pi c v\right)^{1/2}$. The latter sets the transient duration ($\sim$peak time) in the case of spherical ejecta \citep{Arnett79,Arnett80,Arnett82}.
We list this set of variables and definitions in Table~\ref{tab:variables} for reference.

\begin{deluxetable}{lcr}[]
\tablecaption{Main Variables and Definitions\label{tab:variables}}
\tablewidth{0pt} 
\tablehead{ \colhead{notation} & \colhead{definition} & \colhead{interpretation}}
\tabletypesize{\small} 
\startdata 
$R_0$ & $-$ & innermost shell radius at $t=0$\\
$\Delta R_0$ & $^{a}\, \Delta R_{\rm csm}/7$ & postshock shell width at $t=0$\\
$v$ & $-$ & maximum (leading edge) velocity\\
$v_0$ & $^{b}\, 6v_{\rm sh}/7$ & minimum (trailing edge) velocity\\
$M$ & $-$ & CSM shell mass\\
\hline
$\tnot$ & $ \Delta R_0/v$ & shell crossing time\\
$\tdyn$ & $ R_0/v$ & dynamical time\\ 
$\ta$ & $ \left(3\kappa M/4\pi cv\right)^{1/2}$ & ``Arnett'' diffusion timescale\\
$\beta$ & $ v_0/v$ & velocity ratio ($\beta<1$)\\
$E_0$ & $ M v_0^2/2$ & initial internal energy\\
$\tautilde$ & $^{c}\, \sim \tau_0\big/(c/v_0)$ & normalized optical depth (eq.~\ref{eq:tautilde})\\
\hline
$x$ & eq.~(\ref{eq:x}) & dimensionless spatial coordinate\\
$\alpha_n$ & eq.~(\ref{eq:alpha_n}) & solution eigenvalues\\
$f_n(x)$ & eq.~(\ref{eq:f_n}) & spatial eigenfunctions\\
$g_n(t)$ & eq.~(\ref{eq:g_n}) & temporal eigenfunctions\\
\enddata 
\tablenotetext{a}{$\Delta R_{\rm csm} \lesssim R_0$ is the initial (preshock) shell width. It is related to the CSM density near $R_0$ through $\Delta R_{\rm csm} \sim M/4\pi R_0^2 \rho_{\rm csm}$.}
\tablenotetext{b}{$v_{\rm sh}$ is the shock velocity. $v_0$ is also the velocity of the contact discontinuity between CSM and material shocking it (the ejecta).}
\tablenotetext{c}{$\tau_0 \equiv \kappa M/ 4\pi R_0^2$ is the initial optical depth through the CSM.}
\end{deluxetable} 


Before turning our attention to a formal solution of the problem, we conclude this section by deriving an approximate temporal scaling of the light-curve from simple diffusion-time arguments.
The diffusion time of photons from a depth $\Delta r$ within the CSM is $t_{\rm diff} \sim \kappa \rho \Delta r^2/c$. Photons can only effectively escape the medium from within a layer $\Delta r_{\rm diff}$ in which the diffusion time is shorter than the dynamical time, $\Delta r_{\rm diff}(t) \sim \sqrt{c t / \kappa \rho(t)}$.
For a uniform density CSM we can express $\rho(t) \sim M/V(t)$ where $V(t)$ is the CSM volume. The radiated luminosity is roughly equal to the energy that is contained within the diffusion layer divided by the diffusion time. For a uniform energy-density distribution, the former is simply $\sim e(t) \Delta V_{\rm diff}(t)$ where $e(t)$ is the internal energy-density and $\Delta V_{\rm diff}(t) \approx 4\pi R(t)^2 \Delta r_{\rm diff}(t)$ the volume of CSM contained within the diffusion layer (assuming $\Delta r_{\rm diff} \ll R(t)$). Recalling that adiabatic losses cause the energy-density to drop as $e(t) \propto V(t)^{-4/3}$ with time, the total bolometric luminosity prior to the diffusion front crossing the bulk of the CSM (while $\Delta r_{\rm diff} \ll R(t)$; or equivalently $t \ll \ta$) is roughly
\begin{align}
\label{eq:L_diffusion_approximate}
    L(t) 
    &\sim \frac{e(t) \Delta V_{\rm diff}(t)}{t}
    \sim \frac{e_0}{t} \left[\frac{V(t)}{V_0}\right]^{-{4}/{3}} 4\pi R(t)^2 \sqrt{\frac{c t V(t)}{\kappa M}}
    \nonumber \\ 
    &\propto t^{-1/2} R(t)^2 V(t)^{-5/6}
    \propto 
    \begin{cases}
    t^{-4/3} &,\,{\rm planar}
    \\
    t^{-1} &,\,{\rm spherical}
    \end{cases}
\end{align}
where the top case applies in the planar regime $t \ll \tdyn$ in which $R(t)-R(0) \ll R(0)$, and the bottom case for the opposite spherical regime ($t \gg \tdyn$).
As we later show, this approximate solution is in qualitative agreement with the results of our more detailed analysis so long as there exists a regime where $t_0,\tdyn \ll t \ll \ta$ ($\tnot \ll t \ll \tdyn,\ta$) in the spherical (planar) phase.
The detailed solution we derive below (\S\ref{sec:diffusion}) will apply at arbitrary times, and to marginal cases where $\tdyn \sim \ta$ and neither of the regimes in eq.~(\ref{eq:L_diffusion_approximate}) is adequate.

\section{Solution to Diffusion Equation}
\label{sec:diffusion}

The second law of thermodynamics can be written as a partial differential equation for the internal energy density $e(\vec{r},t)$ \citep[e.g.][]{Pinto&Eastman00},
\begin{equation}
\label{eq:second_law}
    \frac{de}{dt} + \frac{4}{3} \frac{\dot{V}}{V} e + \vec{\nabla} \cdot \left( \frac{c}{3 \kappa \rho} \vec{\nabla} e \right) = 0 ,
\end{equation}
where $\kappa$ is the opacity, $\rho(\vec{r},t)$ the density, and $V(t)$ the specific volume. The second term in eq.~(\ref{eq:second_law}) accounts for adiabatic ($PdV$) degradation, while the third term describes radiative losses.

Following \cite{Arnett80}, we make the ansatz that the energy density can be expressed as the product of a spatial function $f(\vec{r})$ and a time-dependent function $g(t)$, such that
\begin{equation}
    e(\vec{r},t) = e_0 f(\vec{r}) g(t) \left[\frac{V(t)}{V_0}\right]^{-4/3}
    .
\end{equation}
Here, $V_0$ is the initial volume, and $e_0$ a normalization factor with units of energy density (such that $f$ and $g$ are dimensionless).
Furthermore, we assume homologous expansion, and rescale the density as
\begin{equation}
    \rho(\vec{r},t) = \rho_0 h(\vec{r}) \left[\frac{V(t)}{V_0}\right]^{-1}
    ,
\end{equation}
with $h$ being another dimensionless function.
With these assumptions, eq.~(\ref{eq:second_law}) can be recast as
\begin{equation}
\label{eq:second_law2}
    \left[\frac{V(t)}{V_0}\right]^{-1} \frac{\dot{g}(t)}{g(t)} = \frac{1}{f(\vec{r})} \vec{\nabla} \left( \frac{c}{3\kappa\rho_0} \frac{\vec{\nabla} f(\vec{r})}{h(\vec{r})}\right)
\end{equation}
in which the time and spatially dependent functions are separated.
This separation of variables allowed \cite{Arnett80} to solve for the eigenfunctions $f$ and $g$ in the simplified case of spherical geometry and constant density ($h=1$). 
Our subsequent derivation differs due to geometric effects that enter through $V(t)$ and the spatial derivatives.

Specifically, the time-dependent volume occupied by the homologously expanding CSM is given by
\begin{align}
\label{eq:V}
    V(t) 
    &= \frac{4\pi}{3} \left[ R(t)^3 - R_{\rm in}(t)^3 \right]
    \\ \nonumber
    &= \frac{4\pi}{3} v^3 \left[ \left(\tdyn+\tnot+t\right)^3 - \left(\tdyn+\beta t\right)^3 \right]
    ,
\end{align}
which describes an expanding spherical shell, rather than a sphere. In particular, in the limit $\tnot+t \ll \tdyn$ we obtain the planar limit in which $V(t) \propto t$. 
We further assume spherical symmetry, such that $f(\vec{r}) = f(r)$ is a function of radius alone.
Finally, a proper choice of coordinates is important for the problem. We define the dimensionless coordinate $x \in [0,1]$ such that {\it at any given time}, $x=0$ corresponds to the interior edge of the expanding CSM, $r=R_{\rm in}(t)$, and $x=1$ to the outer edge at $r=R(t)$. This gives
\begin{equation}
\label{eq:x}
    x \equiv \frac{r-R_{\rm in}(t)}{R(t)-R_{\rm in}(t)}
    .
\end{equation}
This choice of coordinates is dictated by the desire to impose time-independent boundary conditions, and is necessary for proper separation of variables.

Under these assumptions, the energy equation (\ref{eq:second_law2}) can be written as
\begin{align}
\label{eq:second_law3}
    \frac{3\kappa\rho_0}{c} \frac{\left[R(t)-R_{\rm in}(t)\right]^2}{V(t)/V_0} \frac{\dot{g}(t)}{g(t)}
    = 
    &\frac{2}{x+S^{-1}(t)}\frac{f'(x)}{f(x)}
    \nonumber \\ &+
    \frac{f''(x)}{f(x)}
    \approx -\alpha_n
\end{align}
where $S(t) \equiv R(t)/R_{\rm in}(t)-1$ is a measure of the importance of the spherical divergence term. In the planar regime, $S(t) \ll 1$ and the first term on the RHS can be neglected.
In this case, or if $S(t)$ can be considered to vary slowly with time, eq.~(\ref{eq:second_law3}) is spatially and temporally separated so that both sides must be equal to a constant, which we denote $-\alpha_n$. In this regime, the equation is integrable, $\alpha_n$ is an eigenvalue, and we can solve for the eigenfunctions $g_n(t)$ and $f_n(x)$.
We proceed under this assumption, namely that $S^{-1} \gg 1$ and that the resulting planar-regime spatial eigenfunctions reasonably describe the solution. We discuss the applicability of this assumption further in \S\ref{sec:discussion}.

In this case, the spatial eigenfunctions are simply
\begin{equation}
\label{eq:f_n}
    f_n(x) = \sqrt{2} \cos \left( \alpha_n^{1/2} x \right) .
\end{equation}
Imposing a zero-flux boundary condition at the inner edge, $f'(0) = 0$, and the so-called `radiative-zero' condition at the outer boundary, $f(1) = 0$ \citep{Arnett80}, uniquely defines the eigenvalues. They are
\begin{equation}
\label{eq:alpha_n}
    \alpha_n = \left(n-\frac{1}{2}\right)^2 \pi^2 \, , ~~ n=1,2,3...
\end{equation}

The temporal part of the solution is also easily obtained from eq.~(\ref{eq:second_law3}). It is
\begin{align}
\label{eq:g_n}
    g_n(t) 
    &= e^{ -\alpha_n \frac{c}{3\kappa\rho_0V_0} \int \frac{V(t)}{\left[R(t)-R_{\rm in}(t)\right]^2} dt }
    \\ \nonumber
    &= \left[ 1 + \left(1-\beta\right)\frac{t}{\tnot}\right]^{-\alpha_n \left(\frac{\tdyn}{\ta}\right)^2 \frac{\left(1-\beta-\beta \tnot/\tdyn \right)^2}{1-\beta}}
    \\ \nonumber
    &\times e^{ -\alpha_n \frac{t \left[ (1-\beta^3)t + (2-4\beta(\beta+1))\tnot + 6(1-\beta^2)\tdyn \right]}{6(1-\beta)^2 \ta^2} }
    ,
\end{align}
where we recall that $\ta = \left(3 \kappa M / 4\pi c v \right)^{1/2}$ is the characteristic diffusion timescale over which most of the energy is radiated (\citealt{Arnett80}).
The exponential term in $g_n(t)$ is similar to the standard result for an expanding sphere \citep{Arnett80}, however the power-law term is an exclusive feature of the initially planar geometry we consider in the present work.

\subsection{Initial Conditions}

Equations~(\ref{eq:f_n},\ref{eq:alpha_n},\ref{eq:g_n}) define a set of solutions to eq.~(\ref{eq:second_law3}). They form a basis set (delineated by $n$) through which the internal energy density can expressed,
\begin{equation}
\label{eq:e}
    e(x,t) = e_0 \left[\frac{V(t)}{V_0}\right]^{-4/3} \sum_{n=1}^{\infty} e_n f_n(x) g_n(t) .
    .
\end{equation}
The expansion coefficients $e_n$ are set by the initial conditions, $e_n = \langle e(t=0) \vert f_n \rangle$, where we define an inner product $\langle a \vert b \rangle \equiv \int_0^1 a(x) b(x) dx$.
Note that the eigenfunctions $f_n$ (eq.~\ref{eq:f_n}) have been appropriately normalized such that their inner product is $\langle f_n \vert f_m \rangle = 1$ for $n=m$ and $\langle f_n \vert f_m \rangle = 0$ for $n \neq m$.

In the original analyses of \cite{Arnett80,Arnett82}, higher-order $n > 1$ modes were neglected, effectively setting the initial energy density distribution to be $\propto f_1(x)$. \cite{Pinto&Eastman00} pointed out the relevance of higher-order terms in the expansion and discussed the case of uniform energy deposition. In our current model, the case of a fully uniform initial energy distribution, $e(x,t=0) \propto x^0$, would result in a diverging flux at $t = 0$.
This can be understood from the fact that
the diffusion timescale from a given depth $x$ is $t_{\rm diff} \propto (1-x)^2$ in planar geometry, and the corresponding flux is $\propto \int_x^1 e(x) dx/t_{\rm diff}(x)$. This diverges as $x \to 1$ unless the energy density drops to zero near the outer edge.

In the present case of interest, we know that the early light-curve is governed by shock breakout, where radiation from a small region near the CSM edge first manages to escape. This occurs near the luminosity shell, $x = \hat{x}$, where the optical depth is $\tau(>\hat{x}) \approx c/v_{\rm sh}$.
The flux of the breakout emission can be shown to be equivalent to the kinetic shock power per unit area, 
$F_0 = \rho_{\rm csm} v_{\rm sh}^3/2 = (7/6)^3 e_0 v_0$.
Furthermore, this flux can be directly related to the energy density distribution,
\begin{equation}
    F(x,t) = -\frac{c}{3\kappa \rho_0 \Delta R_0} \partial_x e(x,t)
    .
\end{equation}
Assuming a constant initial flux $F(x,t=0) = F_0$ at $x \gtrsim \hat{x}$ implies a linearly decreasing energy density towards $x=1$. This enforces the initial luminosity of our models to be finite, and equal the breakout-flash luminosity.
We further assume that the initial energy density profile at smaller radii is uniform. This is reasonably motivated by the 1D shocked shell structure (the region interior to $x \lesssim \hat{x}$ is adiabatic and has not yet cooled), but this assumption can be easily relaxed.
Combining the above conditions, we adopt an initial energy-density profile
\begin{equation}
\label{eq:e_initial}
    e(x,t=0) = e_0
    \begin{cases}
    1 &, 0 \leq x \leq 1-\tautilde^{-1}
    \\
    \tautilde \left(1-x\right) &, 1-\tautilde^{-1} < x \leq 1
    \end{cases}
\end{equation}
where we have defined $\tautilde \equiv {3\kappa \rho_0 \Delta R_0 F_0}/{e_0 c}$, such that 
\begin{equation}
\label{eq:tautilde}
    \tautilde 
    = \frac{7}{12} \frac{\tau_0}{c/v_{\rm sh}}
    \approx \frac{7^2 \beta}{6^3} \left(\frac{\tdyn}{\ta}\right)^{-2} 
\end{equation}
can be interpreted as the CSM optical depth ($\tau_0 = \kappa \rho_0 \Delta R_0$) normalized by $\sim c/v_{\rm sh}$.
In the final line we have assumed $\rho_0 \approx M/4\pi R_0^2 \Delta R_0$ which is appropriate in the regime $\Delta R_0 \ll R_0$ of interest. 
Note that the transition radius in eq.~(\ref{eq:e_initial}) occurs roughly at the luminosity shell, $1-\tautilde^{-1} \sim \hat{x}$. The requirement that this transition occur at $x > 0$ restricts the allowed parameter-space to $\tautilde > 1$ or $\tdyn/\ta < \sqrt{7^2 \beta / 6^3} \approx 0.5 \beta^{1/2}$. This is equivalent to demanding $\tau_0 > 12c/7v_{\rm sh}$, consistent with the physical requirement that the optical depth exceed $\gtrsim c/v_{\rm sh}$ for radiation to be initially trapped.

Finally, we use the initial condition (\ref{eq:e_initial}) to set the expansion coefficients in eq.~(\ref{eq:e}),
\begin{align}
\label{eq:e_n}
    e_n 
    &= 
    \int_0^1 e(x,t=0) f_n(x) dx
    \nonumber \\
    &= \sqrt{\frac{2}{\alpha_n}} \left(-1\right)^{n+1} \sinc \left( \alpha_n^{1/2} \tautilde^{-1} \right)
    .
\end{align}

\subsection{Emergent Luminosity}

We are now in a position to calculate the bolometric luminosity $L$ that emerges from the outer CSM edge (at $r=R(t)$) by recognizing that
\begin{align}
    L = &4\pi R(t)^2 
    \frac{c}{3\kappa \rho} \left( -\frac{\partial e}{\partial r} \right)_{r=R(t)}
    =
    \frac{4\pi c R_0}{3\kappa \rho_0} 
    \left[\frac{V(t)}{V_0}\right]^{-1/3}
    \nonumber \\ &\times
     \frac{R(t)^2/R_0}{R(t)-R_{\rm in}(t)}
    e_0 \sum_{n=1}^{\infty} e_n g_n(t) \left[ -f'_n(1) \right]
    .
\end{align}
From eqs.~(\ref{eq:f_n},\ref{eq:alpha_n}) we find that 
$f'_n(1) = (-1)^{n} \sqrt{2 \alpha_n}$.
Defining 
\begin{align}
\label{eq:Lambda}
    \Lambda(t) 
    \equiv &\left[\frac{V(t)}{V_0}\right]^{-1/3} \frac{R(t)^2/R_0}{R(t)-R_{\rm in}(t)}
    = \frac{\left( \tdyn + \tnot + t \right)^2}{\tdyn \left[ \tnot + (1-\beta)t \right]}
    \nonumber \\ &\times
    \left[ \frac{\left( \tdyn + \tnot + t \right)^3 - \left( \tdyn + \beta t \right)^3}{\left( \tdyn + \tnot \right)^3 - \tdyn^3} \right]^{-1/3}
\end{align}
we arrive at a final analytic expression for the luminosity,
\begin{align}
\label{eq:L}
    L(t) 
    &= E_0 \frac{\tdyn}{\ta^2} \Lambda(t) \sum_{n=1}^{\infty} (-1)^{n+1} \sqrt{2 \alpha_n} e_n g_n(t)
    \nonumber \\
    &= E_0 \frac{\tdyn}{\ta^2} \Lambda(t) \sum_{n=1}^{\infty} 2 \sinc \left( \alpha_n^{1/2} \tautilde^{-1} \right) g_n(t)
    .
\end{align}
Above, we have defined $E_0 = e_0 V_0$ as the shocked CSM's initial internal energy, and we recall that $\alpha_n$, $g_n(t)$, $e_n$, $\tautilde$, and $\Lambda(t)$ are given by eqs.~(\ref{eq:alpha_n},\ref{eq:g_n},\ref{eq:e_n},\ref{eq:tautilde},\ref{eq:Lambda}), respectively.
For the radiation mediated shock relevant to our work, the initial internal energy satisfies $E_0 = M v_0 ^2/2$.

\section{Properties of the Solution}
\label{sec:properties}

\subsection{Planar vs Spherical Limits}

Let us first examine the light-curve in two limits of interest: the early (planar) regime, applicable while $\tnot + t \ll \tdyn$; and the subsequent spherical regime, at $t \gtrsim \tdyn$. We assume hereafter that the hierarchy of timescales $\tnot < \tdyn < \ta$ is satisfied. The first condition implies an initially thin shell (note that even if $\Delta R_{\rm csm} \sim R_0$, the post shock shell will still be thin, $\Delta R_0/R_0 < 1$). The latter condition, $\tdyn < \ta$, is necessary for establishing the radiation-mediated shock that sweeps the CSM. It is directly related to the requirement that $\tau > c/v_{\rm sh}$ (see eq.~\ref{eq:tautilde}).

\begin{figure}
    \includegraphics[width=0.5\textwidth]{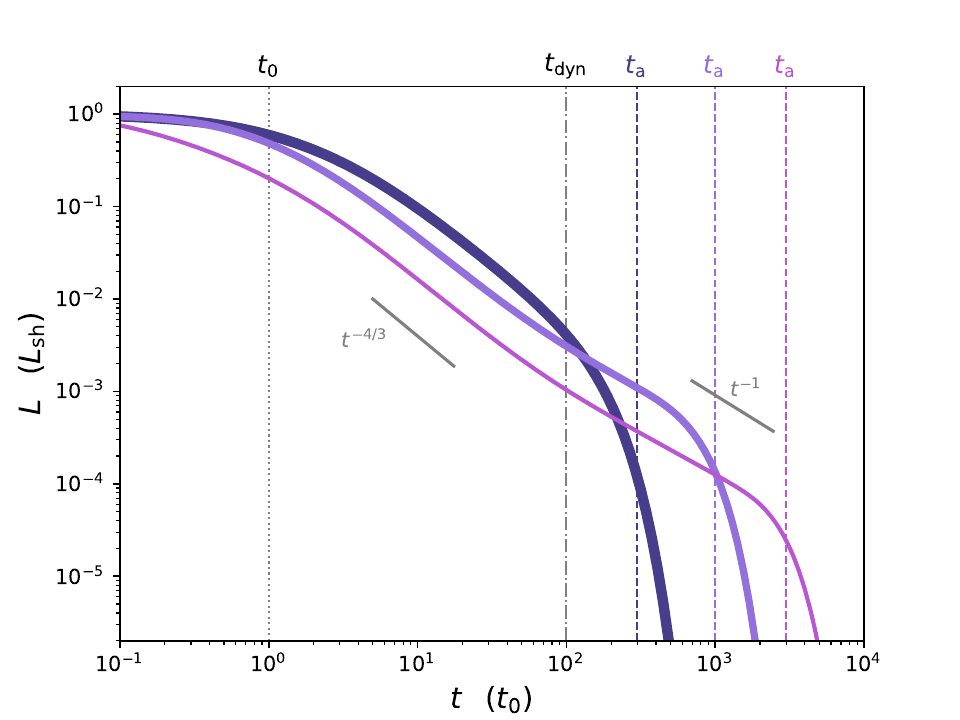}
    \caption{Illustrative light-curves: the light-curve morphology depends on the timescales $\tdyn$, $\ta$, in relation to the shell width crossing time $\tnot$ (note that the hierarchy $\tnot < \tdyn < \ta$ is always satisfied). For a fiducial velocity ratio $\beta = 0.5$, and $\tdyn/\tnot = 100$, different colored curves show light-curves for increasing values of $\ta/\tdyn = 3,10,30$ (dark to light; equivalent to successively larger optical depths $\tautilde \approx 1,10,100$, eq.~\ref{eq:tautilde}).
    The luminosity is normalized by the initial (shock) luminosity $L_{\rm sh}$ (eq.~\ref{eq:L_0}), and time is plotted in units of the initial shell-crossing time $\tnot$ (Tab.~\ref{tab:variables}).
    The light-curve drops as $\sim t^{-4/3}$ during the planar phase, continues to drop (at a slower rate, roughly $L \propto t^{-1}$) while $\tdyn < t < \ta$, and declines supra-exponentially at $t \gtrsim \ta$. See \S\ref{sec:properties} for further details.}
    \label{fig:LC_shape}
\end{figure}

The function $\Lambda(t)$ that enters the luminosity expression (eqs.~\ref{eq:Lambda},\ref{eq:L}) can be approximated as
\begin{equation}
    \Lambda(t) \approx
    \begin{cases}
    \frac{\tdyn}{\tnot} \left[ 1 + (1-\beta) \frac{t}{\tnot} \right]^{-4/3}
    &, \tnot + t \ll \tdyn
    \\
    \left[\frac{3\tnot}{(1-\beta)^2\tdyn}\right]^{1/3}
    &, t \gg \tdyn
    \end{cases}
\end{equation}
in the planar (top) and spherical (bottom) regimes. 
Using this result, and returning explicit expressions for the expansion coefficients $e_n$ (eq.~\ref{eq:e_n}) into eq.~(\ref{eq:L}), we find that the luminosity in the planar regime is
\begin{align}
\label{eq:L_planar}
    L\left(t \ll \tdyn\right) \approx
    \frac{7^2}{6^3} \frac{\beta E_0}{\tnot}
    &\sum_{n=1}^\infty 2\alpha_n^{-1/2} \sin \left( \alpha_n^{1/2} \tautilde^{-1} \right)
    \nonumber \\
    &\times 
    \left[ 1 + (1-\beta) \frac{t}{\tnot}\right]^{-p_n}
\end{align}
where
\begin{equation}
\label{eq:pn}
    p_n \equiv \frac{4}{3} + \alpha_n \left(\frac{\tdyn}{\ta}\right)^2 \frac{(1-\beta-\beta \tnot/\tdyn)^2}{1-\beta}
    .
\end{equation}
This shows that, after an initial shell crossing time $\tnot$, the planar light-curve drops as a power-law in time $L \propto t^{-p_n}$. For $\tdyn \ll \ta$, the exponent of low order modes (small $\alpha_n$) is dominated by the first term, and one recovers the familiar $L \propto t^{-4/3}$ scaling that is implied by eq.~(\ref{eq:L_diffusion_approximate}). Physically, this scaling describes the planar luminosity in the limit of large optical depth, where only adiabatic losses are important.

A novel feature of the model derived in the present work is that the light-curve of individual modes may decay more rapidly than this $t^{-4/3}$ scaling.
This occurs because our model allows for the possibility that radiative losses may be important already during the planar phase. Accordingly, this steepening is inversely proportional to the CSM optical depth (eq.~\ref{eq:tautilde}),
\begin{equation}
\label{eq:Delta_pn}
    \Delta p_n 
    \equiv p_n - \frac{4}{3}
    \approx \alpha_n \left(\frac{\tdyn}{\ta}\right)^2 (1-\beta)
    \propto \tautilde^{-1} .
\end{equation}
This expression shows that the decay rate of the dominant, lowest-order, mode can be steepened by as much as $\Delta p_1 \approx 0.14 \tautilde^{-1}$. Higher order modes will be increasingly dramatically impacted (e.g. $\Delta p_2 \lesssim 1.3 \tautilde^{-1}$, $\Delta p_3 \lesssim 3.5 \tautilde^{-1}$, etc).
This steepening is most relevant for CSM with ``marginal'' optical depth, $\tautilde \gtrsim 1$.\footnote{We consider this case ``marginal'' in the sense that if $\tau_0$ were smaller than $c/v_{\rm sh}$ (i.e. $\tautilde < 7/12 \simeq 0.6$) then a radiation-mediated shock is never formed, and the physical situation differs dramatically from that discussed in the present work (see e.g. \citealt{Margalit+22}).}
Intriguingly, numerical work has recently found that precisely such configurations may be necessary for modeling at least some well-observed FBOTs \citep{Rest+18}.
In the limit $\tdyn \ll \ta$ when the optical depth is large, then $\Delta p_n \ll 1$ and this term introduces a weak logarithmic correction $\propto 1 - \Delta p_n \ln \left[ 1 + (1-\beta)t/\tnot \right] + \mathcal{O}\left(\Delta p_n^2 \right)$ to the standard $\sim t^{-4/3}$ scaling of the luminosity.
Logarithmic corrections of a similar form have also been recently studied by \cite{Faran&Sari19}.

We can also verify from eq.~(\ref{eq:L_planar}) that the luminosity at $t \ll \tnot$ reduces to the breakout luminosity.
Noting that 
$\sum_{n=1}^\infty 2\alpha_n^{-1/2} \sin ( \alpha_n^{1/2} \tautilde^{-1} ) = 1$ for any $\tautilde \geq 1$, 
the initial luminosity is simply
\begin{equation}
\label{eq:L_0}
    L(t=0) =
    \frac{7^2}{6^3} \frac{\beta E_0}{\tnot}
    = \frac{M v_{\rm sh}^3}{2\Delta R_{\rm csm}}
    = L_{\rm sh} ,
\end{equation}
where 
$L_{\rm sh} = 2\pi R_0^2 \rho_{\rm csm} v_{\rm sh}^3$ is the kinetic shock power. This is the expected result, since the initial (breakout) luminosity must equal the shock power.

Turning now to the spherical regime, we find that the luminosity (eq.~\ref{eq:L}) in this phase is
\begin{align}
\label{eq:L_spherical}
    L\left(t \gg \tdyn\right) 
    &\approx
    \frac{7^2}{6^3} \frac{\beta E_0}{\tnot} \frac{3^{1/3}}{(1-\beta)(1-\beta^3)^{1/3}} 
    \\ \nonumber
    &\times 
    \left(\frac{\tdyn}{\tnot}\right)^{-4/3}
    \sum_{n=1}^\infty 2\alpha_n^{-1/2} \sin \left( \alpha_n^{1/2} \tautilde^{-1} \right)
    \\ \nonumber
    &\times 
    \left[(1-\beta)\frac{t}{\tnot}\right]^{-\Delta p_n} 
    e^{-\alpha_n \left(\frac{t}{\ta}\right)^2 \frac{(1-\beta^3)}{6(1-\beta)^2}}
    .
\end{align}
Here, the light-curve of any individual mode is proportional to $t^{-\Delta p_n}$ (eq.~\ref{eq:Delta_pn}) times a Gaussian term $\sim e^{-\alpha_n (t/\ta)^2}$. The light-curve shape then depends on the ratio of dynamical and diffusion timescales.
If $\ta \gg \tdyn$ then the power-law term has a very weak time dependence and the light-curve is dominated by the exponential. In this regime the luminosity of individual modes is $\sim$flat while $\tdyn \ll t < \ta/\alpha_n^{1/2}$, and drops dramatically at $t \gtrsim \ta/\alpha_n^{1/2}$. This behavior is qualitatively consistent with the standard \cite{Arnett80} model that was derived in the spherical regime alone.
As discussed further in \S\ref{sec:modes}, the initial conditions imply that many modes must be retained in the series expansion, and the combination of these modes yields a light-curve that differs dramatically from the description above.
Additionally, if $\ta$ is only slightly larger than $\tdyn$ (marginal optical depth) then the power-law decay continues to be important also during the spherical phase.

Additional insight into the effect of the $\propto t^{-\Delta p_n}$ term, a novel feature of our model, on the spherical-phase light-curve can be gained by considering the impact of this term between time $\tdyn$ and $\ta/\alpha_n^{1/2}$.
The power-law term contributes to a drop in the light-curve of a given mode by a factor 
\begin{align}
\label{eq:pwrlaw_spherical_drop}
    \delta_n 
    &\equiv 
    \left(\frac{\tdyn}{\ta/\alpha_n^{1/2}}\right)^{\alpha_n \left(\frac{\tdyn}{\ta}\right)^2 (1-\beta)}
    \\ \nonumber
    &\underset{\tdyn \ll \ta/\alpha_n^{1/2}}{\approx} 1 + (1-\beta) \left(\frac{\tdyn}{\ta}\right)^2 \ln \left(\frac{\tdyn}{\ta}\right) + ...
\end{align}
between $t=\tdyn$ and $t=\ta/\alpha_n^{1/2}$. 
If $\tdyn \ll \ta$ then this term is $\delta_n \approx 1$, with only a small logarithmic correction. This is consistent with the expectation that planar radiative losses should not impact the light-curve deep within the spherical regime.
Even if $\tdyn \lesssim \ta$,
the power-law term will only have a minor impact during the spherical regime (though it may have a substantial effect during the planar regime). This is because there is only a small dynamical range of time $\tdyn \lesssim t \lesssim \ta/\alpha_n^{1/2}$ before the Gaussian term takes over and the light-curve drops supra-exponentially. Equation~(\ref{eq:pwrlaw_spherical_drop}) formally attains a minimum at $\ta \alpha_n^{-1/2}/\tdyn = \sqrt{e}$
for which 
$\delta_n = e^{-(1-\beta)/2e} \gtrsim 0.9$. Therefore the power-law term has, at most, a $\lesssim 10\%$ effect on the light-curve within the spherical regime ($t > \tdyn$) and this effect vanishes for $\ta \gg \tdyn$.
This implies that $\delta_n \approx 1$ and that the light-curve {\it of a given eigenmode} scales as $\sim e^{-\alpha_n (t/\ta)^2}$ within the spherical regime, $t \gg \tdyn$. We point out once again that this is the same qualitative temporal scaling that is derived from consideration of a purely spherical problem \citep[e.g.][]{Arnett80}.

\subsection{High-Order Modes}
\label{sec:modes}

After deriving approximate expressions for the light-curve during the planar/spherical regimes and discussing these solutions from the point of view of a single term in the eigenmode expansion, we now turn to examine the full light-curve that results from summing different eigenmode contributions.
Fig.~\ref{fig:LC_shape} shows such light-curves for different parameter choices. The light-curve shape depends on the ratio of characteristic timescales $\tnot$, $\tdyn$, and $\ta$, and on the velocity ratio $\beta$ between innermost and outermost material in the expanding shell. A fifth parameter, the total energy $E_0$, is only relevant for the normalization (recall that these parameters are directly related to the physical CSM properties: $M$, $R_0$, $\Delta R_{\rm csm}$, $v_0$, and $v$; see Tab.~\ref{tab:variables}).

To better compare the light-curve morphology, we plot the luminosities in Fig~\ref{fig:LC_shape} normalized by the initial (shock) luminosity, $L_{\rm sh}$ (eq.~\ref{eq:L_0}). We similarly choose to plot the time axis in units of $\tnot$ (dotted grey curve).
In this normalized phase space, the light-curve depends entirely on the values of $\tdyn/\tnot$, $\ta/\tdyn$, and $\beta$. We choose a fiducial $\beta = 0.5$, and take $\tdyn/\tnot = 100$ (dot-dashed grey vertical curve), however the results are qualitatively similar for other physical choices of these parameters.
Light-curves for varying values of $\ta$ are shown as different colored curves, with $\ta/\tdyn = 3, 10, 30$ (dark to light colors). This corresponds to increasing initial CSM optical depths: $\tautilde \approx 1, 10, 100$, respectively (eq.~\ref{eq:tautilde}).

As discussed in the previous subsection, the light-curves attain an initial value $L(t\ll\tnot)=L_{\rm sh}$ and falloff as roughly $\sim t^{-4/3}$ during the planar regime, $\tnot \ll t \lesssim \tdyn$.
However, the total light-curve differs appreciably from the previously-discussed single-mode light-curve at times $\tdyn \ll t < \ta$. In this regime we observe that, in the limit $\ta \gg \tdyn$, the light-curve scales approximately as $L \propto t^{-1}$, consistent with eq.~(\ref{eq:L_diffusion_approximate}). This is opposed to the ``flat'' 
light-curve that any single mode contributes to eq.~(\ref{eq:L_spherical}), and is a consequence of the interplay of different modes.
We return to this point soon.

\begin{figure}
    \centering
    \includegraphics[width=0.5\textwidth]{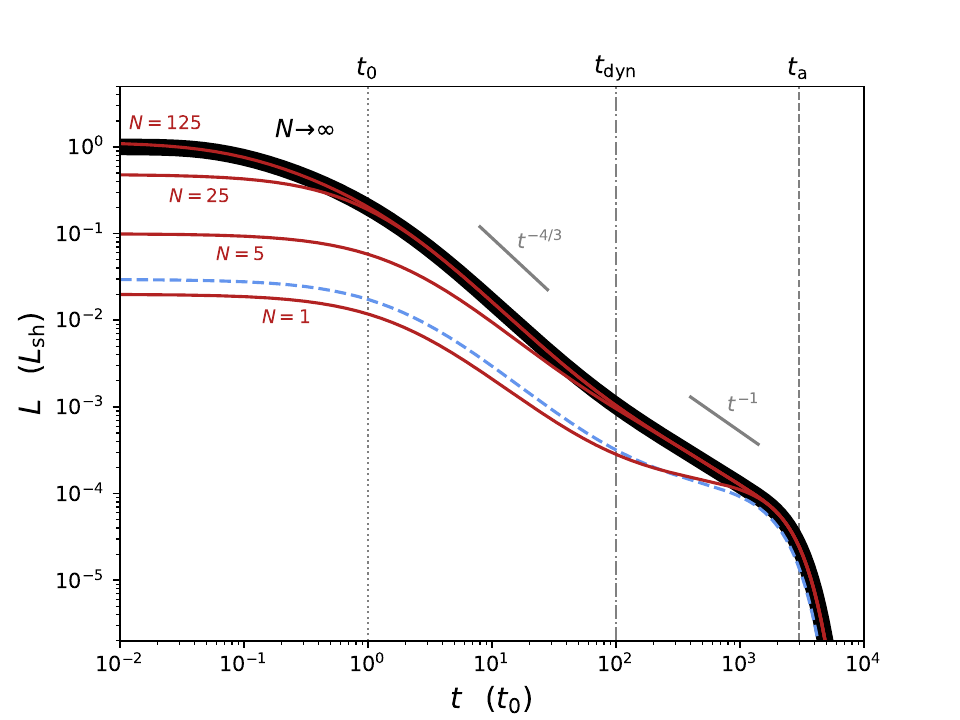}
    \caption{Relevance of higher-order eigenmodes in the series expansion: normalized light-curve for fiducial parameters $\tdyn = 100\tnot$, $\ta = 30\tdyn$, and $\beta=0.5$ is shown in solid black. Thin red curves show the respective light-curves that result if only a small number of terms $n=1,...,N$ is retained in the series expansion (eq.~\ref{eq:L}). For $\ta \gg \tdyn$ ($\tautilde \gg 1$; eq.~\ref{eq:tautilde}) the light-curve requires $N \sim \mathcal{O}(\tautilde)$ terms to converge, and is not well-described at early times by the lowest-order $n=1$ eigenmode ($\tautilde \approx 100$ in the illustrated case). CSM configurations with $\ta \gtrsim \tdyn$ (not shown) are more reasonably approximated by the first eigenmode. Also shown is the light-curve of a one-zone model (dashed-blue; eq.~\ref{eq:Appendix_OneZone}), which closely follows the $n=1$ mode. See \S\ref{sec:modes} for further details.}
    \label{fig:modes}
\end{figure}

The importance of high-order modes can also be examined by comparing light-curves that retain fewer terms in the series eigenvalue expansion. This is shown in Fig.~\ref{fig:modes}, where a representative light-curve is compared to the respective light-curves that result if only $n = 1,..,N$ terms are retained in the expansion.
For the chosen parameters, the fundamental mode (curve labeled $N=1$) provides a poor description of the full light-curve (black) at early times, $t \ll \ta$.
As the total number of modes $N$ is increased, the light-curve becomes accurate at successively earlier epochs. Overall, we find that $N \sim \mathcal{O}(\tautilde)$ terms are necessary before the light-curve converges to its asymptotic value (for the parameters illustrated in Fig.~\ref{fig:modes}, $\tautilde \approx 100$).

The fact that larger CSM optical depth (larger $\ta/\tdyn$) requires a greater number of eigenmodes is directly related to the fact that shock breakout occurs closer to the CSM outer edge if $\ta \gg \tdyn$ (eqs.~\ref{eq:e_initial},\ref{eq:tautilde}). This implies a larger gradient of the initial energy-density distribution (eq.~\ref{eq:e_initial}) that can only adequately be described by high-$n$ modes. 
Specifically, resolving the breakout luminosity-shell lengthscale $\sim (1-\hat{x}) \approx \tautilde^{-1} \ll 1$ with the eigenfunctions $f_n(x)$ (eq.~\ref{eq:f_n}) requires mode numbers for which $\alpha_n^{1/2} (1-\hat{x}) \sim \pi/2$, or $n \sim \tautilde$.
Increasing power is therefore placed into higher order eigenmodes as $\tautilde \gg 1$.
An alternative way to view this is by examining eq.~(\ref{eq:L}). The luminosity expansion coefficients are proportional to $\sinc (\alpha_n^{1/2} \tautilde^{-1})$. This is dominated by terms with $\alpha_n^{1/2} \tautilde^{-1} \lesssim \pi/2$ because the $\sinc$ function yields cancelling oscillatory terms of decreasing amplitude for larger arguments (higher $n$), therefore these do not contribute appreciably to the total luminosity.
This approach also allows us to understand the $L \sim t^{-1}$ scaling of the light-curve at times $\tdyn \ll t < \ta$.

In the limit where there exists a time span such that $\tdyn \ll t \ll \ta$, each relevant mode contributes $\sim$equally to the light-curve. This is because, as discussed above, most of the power is in modes with $\alpha_n^{1/2} \tautilde^{-1} \lesssim 1$.
If we define the ``lifetime'' of a given mode as $\ta^n \equiv \ta/\alpha_n^{1/2}$ (at times $t \gtrsim \ta^n$ the contribution of the $n^{\rm th}$ mode is negligible due to the supra-exponential decline) then at times $t \gg \ta \tautilde^{-1}$ only the subset of modes with $\alpha_n^{1/2} \tautilde^{-1} \ll 1$ contribute to the light-curve. In this regime the expansion coefficients are $\sinc (\alpha_n^{1/2} \tautilde^{-1}) \approx 1$, independent of $n$. The light-curve therefore declines by a factor $\delta L_n \sim 1-1/n$ over a fractional change in time $\delta t_n \sim \ta^{n-1}/\ta^n$ (over which the $n^{\rm th}$ mode ``dies out''). This yields a light-curve decline-rate 
\begin{align}
\label{eq:dlnLdlnt}
    \left.\frac{d\ln L}{d\ln t}\right\vert_{\ta^n}
    &\sim \frac{\ln(\delta L_n)}{\ln(\delta t_n)} \sim \frac{\ln(1-1/n)}{\ln\left[({2n-1})/({2n-3})\right]}
    \\ \nonumber
    &\underset{n \gg 1}{\approx} -1 + \frac{1}{2n} 
    +...
    \approx -1 + \frac{\pi}{2} \left(\frac{t}{\ta}\right)
    +...
\end{align}
that is close to the observed $L \propto t^{-1}$, or slightly shallower (Fig.~\ref{fig:LC_shape}).
The expression above is evaluated at $t = \ta^n$, which has allowed us to relate the mode number to time, $n \sim {1}/{2}+({t}/{\ta})^{-1}/\pi$.
Note also that eq.~(\ref{eq:dlnLdlnt}) is only valid in the regime $\tdyn, \ta \tautilde^{-1} \ll t \ll \ta$.

Eq.~(\ref{eq:dlnLdlnt}) shows that the $L \propto t^{-1}$ decay within the spherical regime is a consequence of the initial conditions that we have imposed (eq.~\ref{eq:e_initial}). That is---even though the light-curve of any given mode within our model is qualitatively similar to the classical \cite{Arnett80} result---the ensemble of modes that is governed by the initial conditions appropriate for the shocked-CSM scenario implies a light-curve that differs substantially from the \cite{Arnett80} result at times $t \ll \ta$.
This is also consistent with the simple diffusion-time estimate (eq.~\ref{eq:L_diffusion_approximate}), that is similarly governed by the uniform density/energy-density initial conditions.

As a final point we also plot in Fig.~\ref{fig:modes} the light-curve that results from a ``one-zone model'' (dashed blue curve). This model is described in Appendix~\ref{sec:one-zone}, and can be viewed as an extension of spherical one-zone models \citep[e.g.][]{Kasen&Bildsten10,Piro15} to our current geometry. This approach neglects spatial aspects of the problem, producing a light-curve that is qualitatively similar to that of a single eigenmode. It is therefore not surprising that the dashed-blue curves in Fig.~\ref{fig:modes} closely follow the $N=1$ (single, lowest-order eigenmode) light-curve. Clearly, this approach is limited if higher-order modes are important to the total light-curve (as they generally are for $\tautilde \gg 1$).
A second, albeit more minor, deficiency of the one-zone model, even in cases where the $n=1$ mode dominates the light-curve, is the ``wrong'' prefactor that enters the temporal evolution. This arises because the one-zone approach has no information regarding the spatial eigenvalues. This point is discussed in greater detail in \cite{Khatami&Kasen19} in the context of typical SNe. It is purely a coincidence that in our present scenario the one-zone model predicts a prefactor~$=3$ that is similar to the lowest-order eigenvalue $\alpha_1 = 2.47$ (eq.~\ref{eq:alpha_n}), and that therefore this model reasonably approximates the lowest order mode. In the spherical case there is a larger discrepancy between the proper eigenvalue and one-zone model predictions.

\section{Observational Consequences}
\label{sec:observations}

In the previous sections we have derived and discussed solutions of the diffusion equation for a dense, shocked shell of expanding material. The primary result is an analytic expression for the bolometric luminosity of the shell as a function of time (eq.~\ref{eq:L}).
This situation may be relevant to different astrophysical settings, and in particular the scenario in which the shocked material is an extended optically thick CSM that is shocked by a SN ejecta. This scenario is a leading model in explaining the emerging class of fast optical transients, and we discuss below observational implications of our model in this context.

\begin{figure}
    \centering
    \includegraphics[width=0.5\textwidth]{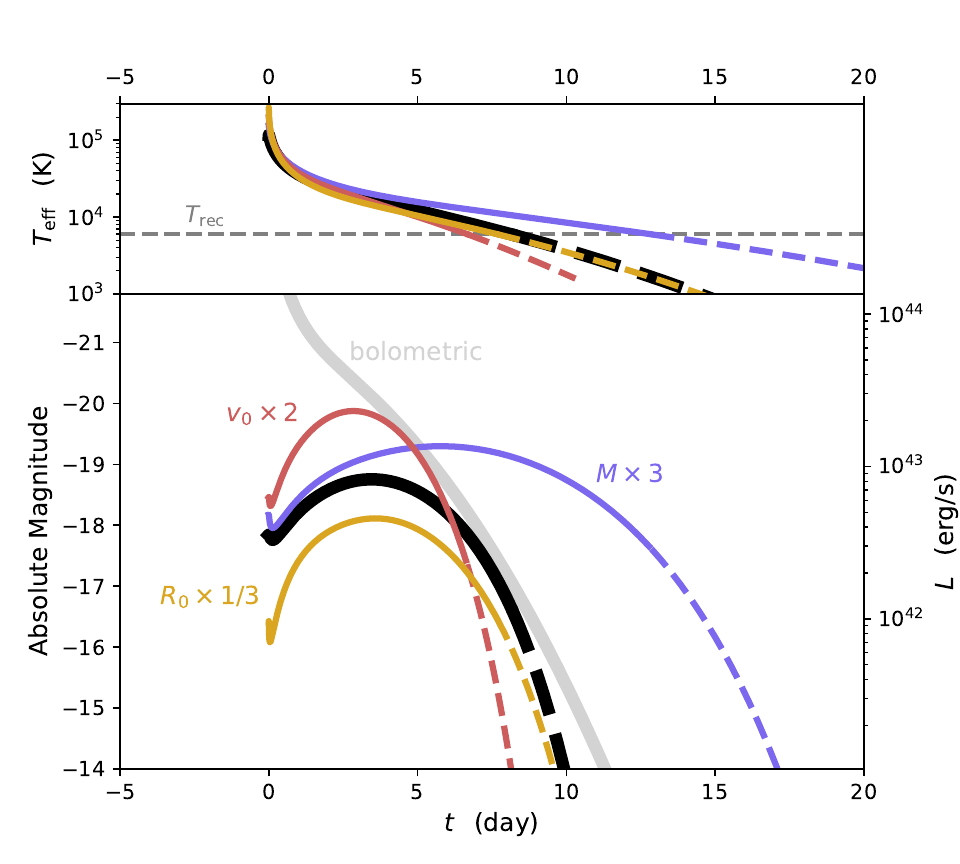}
    \caption{Optical light-curve of a fiducial CSM shell of mass $M = 0.1M_\odot$ at radius $R_0 = 10^{14}\,{\rm cm}$, (post-shock) width $\Delta R_0 = R_0/7$, that is shocked by an outflow with velocity $v_0 = 10^9\,{\rm cm \, s}^{-1}$, and assuming $\beta = 0.5$ (black). The optical luminosity is estimated as $\nu L_\nu$ at a wavelength of $5000 \, \Angstrom$ assuming a blackbody SED at temperature $T_{\rm eff}$ (shown in top panel), and an opacity $\kappa = 0.34 \, {\rm cm}^2 \, {\rm g}^{-1}$. The bolometic luminosity is shown in solid grey, while the red, blue and yellow curves show the effect of varying different source parameters (labeled). The effective temperature drops below the H-recombination temperature, $T_{\rm rec} \approx 6000\,{\rm K}$, on timescales $\gtrsim$a few times optical peak. Dashed curves show this portion of the evolution.}
    \label{fig:optical}
\end{figure}

CSM properties that are inferred from observations of fast transients differ between events (and based on modelling methods), but typical quoted values span $M \sim 10^{-3}-10^{-1} M_\odot$, $R_0 \sim 10^{13}-10^{15} \, {\rm cm}$, and $v_0 \sim 10^4 \,{\rm km \, s}^{-1}$.
Our model is a function of these three physical variables ($M$, $R_0$, and $v_0$) and two dimensionless parameters whose fiducial values we take to be $\beta = 0.5$, $\Delta R_0 / R_0 = 1/7$ (see Table~\ref{tab:variables}). Sufficiently well-sampled bolometric light-curves can, in principle, constrain all five of these parameters. In practice however, band-limited and/or sparsely sampled light-curves would likely limit such constraints. We leave detailed comparison and fitting of our model to observed data for future work and in the following focus primarily on comparing our results to previous models.

In Fig.~\ref{fig:optical} we show example light-curves predicted by our model for several different choices of these physical parameters.
To better compare these results to observed transients, one must consider the fact that most of the early-time bolometric luminosity is emitted in the UV, where observations are typically sparse. The optical light-curve will therefore deviate substantially from the bolometric (eq.~\ref{eq:L}).
To crudely estimate the optical-band luminosity, we assume a blackbody spectrum characterized by temperature $T_{\rm eff} = \left[ L(t) / 4\pi \sigma_{\rm SB} R_{\rm ph}(t)^2 \right]^{1/4}$, where $\sigma_{\rm SB}$ is the Stefan-Boltzmann constant
and $R_{\rm ph}(t) = R(t) - 2 V(t) / 3 \kappa M$ is the photospheric radius (at which $\tau = 2/3$; eq.~\ref{eq:V}).\footnote{Note that once $R_{\rm ph}(t) < R_{\rm in}(t) = v (\tdyn+\beta t)$ the CSM enters the nebular phase and the solutions derived in this work are no longer applicable (more specifically, radiation becomes free-streaming and the diffusion approximation [eq.~\ref{eq:second_law}] breaks down).}
Fig.~\ref{fig:optical} shows the optical luminosity of a fiducial CSM shell,
estimated as $\nu L_\nu$
at a wavelength of $5000 \,\Angstrom$ (black). The optical light-curve has a clear peak at $\sim 3 \,{\rm day}$ even though the bolometric luminosity (light-grey curve) monotonically declines.
This is caused by the rapidly declining temperature (top panel). Therefore, within this model: {\it the optical light-curve maxima occurs when the temperature drops through the optical band}.
This is qualitatively distinct from $^{56}$Ni-powered core-collapse or thermonuclear SNe, whose optical peak is set by photon diffusion time \citep{Arnett80,Arnett82,Pinto&Eastman00}.
This situation implies a strongly chromatic dependence of optical peak, with bluer bands peaking earlier.

Note that our modeling does not capture the physics that governs the rising part of the bolometric luminosity. This is set by (the shorter of) the shell light-crossing time, $\sim R_0/c \ll \, {\rm day}$, or diffusion of photons during the initial shock-breakout timescale, $t_{\rm bo} \sim c/\kappa \rho_{\rm csm} v_{\rm sh}^2$.
Since both these timescales are typically $\ll$ than the time of optical peak, this does not affect our conclusion regarding the optical-band light-curve peak (but see e.g. \citealt{Katz+12} for detailed calculations of light-crossing time effects).

Fig.~\ref{fig:optical} also shows the effective temperature evolution. On timescales $\gtrsim$a few times optical peak, the temperature drops below $T_{\rm rec} \sim 6000\,{\rm K}$ and hydrogen may begin to recombine. The effect of H-recombination on the resulting light-curve is not captured by our present model, so the light-curves in Fig~\ref{fig:optical} may be unreliable after this time. This is indicated by the dashed portion of the curves.
Because recombination generally occurs at times $t_i > \ta$ for CSM configurations of interest, this effect may not appreciably impact the resulting bolometric luminosity \citep{Popov93}. The photospheric temperature and resulting optical-band light-curve will still change even if the bolometric luminosity does not ($T_{\rm eff}$ will be fixed to $T_{\rm rec}$, and the optical-band light-curve will closely track the bolometric). We postpone more detailed investigation of recombination in the context of shock-cooling CSM to future work, and refer to e.g. \cite{Faran+19} for a recent discussion in the context of IIP SNe.

\subsection{Comparison with Previous Work}

We conclude this section by comparing our results to previous modeling of dense CSM shock-cooling.
The work of \cite{Piro15} is particularly notable in this context as it has been used to interpret several observed transients \citep[e.g.][]{Ho+19,Yao+20}.
\cite{Piro15} extended previous peak-timescale/luminosity estimates \citep[e.g.][]{Chevalier&Irwin11} by providing a time-dependent light-curve that could be fit to data. This was derived from a one-zone spherical model whose result is analogous to that of \cite{Arnett80}.\footnote{Formally, the only difference between these models is the normalization of $\ta$ (labeled $t_{\rm p}$ in \citealt{Piro15}): \cite{Arnett80} adopts a prefactor determined by the lowest-order eigenvalue of the spatially-dependent (spherical) problem, while \cite{Piro15} normalizes this quantity based on results from simulations.}
We also compare our results to the recent \cite{Piro+21} model that extended \cite{Piro15} with an updated prescription for the early light-curve.
Our current work differs from \cite{Piro+21} in two major respects: 
(i) the assumed shocked-CSM density distribution, and (ii) the treatment of radiative diffusion. 
\cite{Piro+21} assume that the shocked-CSM density profile is a broken power law with an outer index $\sim 10$ and inner index $\sim 1$. Such a profile is motivated by e.g. \cite{Chevalier&Soker89}, and describes well homologous expansion of material following shock breakout from a star, where the outer density profile follows $\rho(r) \propto \left( R_0-r \right)^n$ with $n \sim 1.5-3$ (see \S\ref{sec:description} for further details). Here we focus instead on the case where the CSM density profile falls abruptly at $r \sim R_0$. This implies a post-shock density profile that has only a ``flat'' interior (the mass in the outer steep component is negligible; this is equivalent to taking the outer power-law index $\to \infty$).
Furthermore, the treatment of radiative diffusion differs between out current work and \cite{Piro+21}. Here we explicitly solve the radiative diffusion equation accounting for energy transport and radiative losses as a function of spatial position within the CSM. \cite{Piro+21} adopt an approximate approach analogous to \cite{Nakar&Sari10} that is roughly equivalent as long as the diffusion wave is traversing the steep outer density profile (which is effectively non-existent in our scenario). At later times, when the diffusion front enters the ``flat'' inner density-profile where the bulk of material is contained, \cite{Piro+21} adopt a one-zone approach similar to \cite{Piro15}. This neglects the spatial energy-density distribution that is treated in detail in our current work.

\begin{figure}
    \centering
    \includegraphics[width=0.5\textwidth]{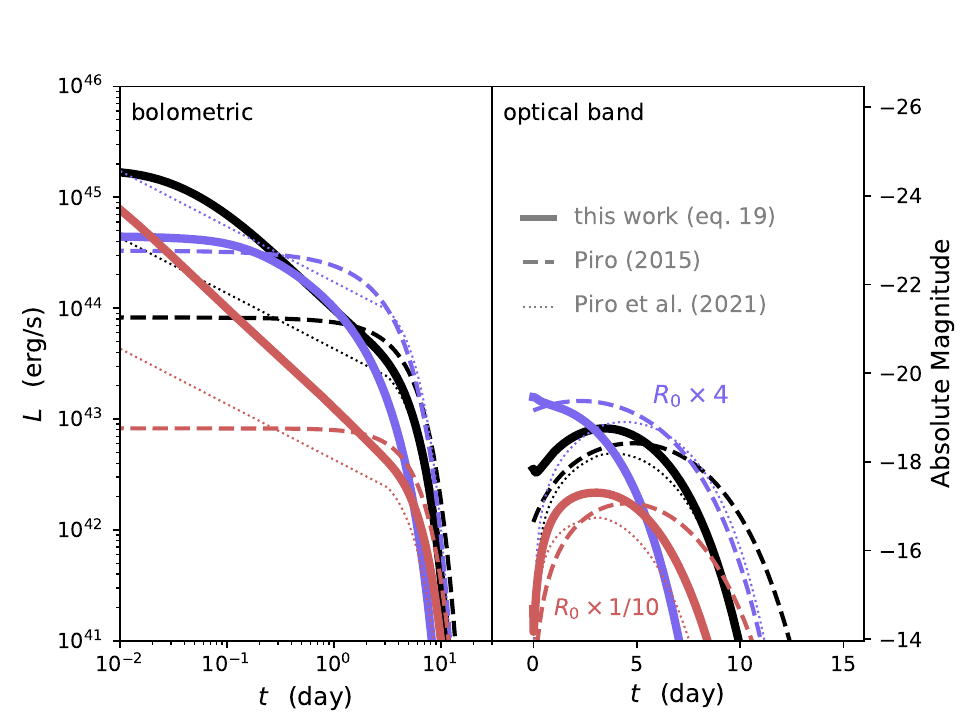}
    \caption{Comparison of the model derived in our present work (solid curves; eq.~\ref{eq:L}) with the shock cooling model of \cite{Piro15} and \cite{Piro+21} (dashed and dotted curves, respectively). We show the bolometric luminosity (left panel) and optical-band luminosity (right; $\nu L_\nu$ at $5000 \Angstrom$) for a fiducial model (black) with the same CSM parameters as in Fig.~\ref{fig:optical}. Blue (red) sets of curves show the light-curves of CSM with similar parameters, except that the radius $R_0$ is changed to be four times larger (ten time smaller). The bolometric light-curves predicted by the \cite{Piro15} one-zone model and the \cite{Piro+21} model differ substantially from eq.~(\ref{eq:L}), particularly at early times. The optical band light-curves are qualitatively similar, but differ at the detailed level.}
    \label{fig:Piro_Comparison}
\end{figure}

In Figure~\ref{fig:Piro_Comparison} we compare the light-curves resulting from our present model (eq.~\ref{eq:L}; solid curves) to those predicted by \cite{Piro15} and \cite{Piro+21} (dashed and dotted curves, respectively) for the same CSM configurations (different colors). On the left we show the bolometric luminosity in log-space, while the right-hand panel shows the optical-band light-curve calculated at $5000\,\Angstrom$.
As in Fig~\ref{fig:optical}, our fiducial model assumes $M = 0.1M_\odot$, $R_0 = 10^{14}\,{\rm cm}$, $\Delta R_0 = R_0/7$, $v_0 = 10^9\,{\rm cm \, s}^{-1}$, and $\beta = 0.5$ (black curves).
We apply the same parameters to the \cite{Piro15,Piro+21} models, where the only ambiguity relates to the fact that we have two velocity scales in our present work ($v_0$ and $v = v_0/\beta$) while there is only one in the aforementioned models. In Fig~\ref{fig:Piro_Comparison} we take this velocity ($v_{\rm e}$ in the notation of \citealt{Piro15}) to be $v_0$, comparable to the shock velocity. 
This ensures that the initial thermal energy $E_0$ is the same throughout all models. 
An alternative choice of $v_{\rm e} = v$ would result in more luminous light-curves that rise and fall on shorter timescales, but that still do not quantitatively match the results of our current model (in particular, the peak luminosity would be over-predicted).\footnote{To facilitate comparison with \cite{Piro15} we explicitly use velocity, rather than energy, in the definition of $t_{\rm p}$: specifically we calculate $t_{\rm p}$ in the \cite{Piro15} model using his eq.~(4) multiplied by $\simeq 0.231$, as implied by eq.~(6) of that work. 
For the \cite{Piro+21} model we take the kinetic energy to be $E_e = M v_0^2/2$ and calculate the characteristic velocity using eq.~(4) of that work.
}

The optical-band light-curves of both models agree at a qualitative level, but there are clear differences in the shape and rise/fall timescales.
For example, we find that the fade times implied by our current model are $t_{1/2} = 2.9, 3.1, 2.8$ day for the fiducial, $R_0 \times 4$, and $R_0 \times 1/10$ models, respectively. Here we have defined $t_{1/2}$ as the time elapsed between optical peak and till the light-curve drops to half optical maximum. The \cite{Piro15} model instead implies $t_{1/2} = 3.7, 4.1$, and $3.2$ day for these same models,
whereas the \cite{Piro+21} model implies fade times that are closer to our current results: $t_{1/2} = 3.0, 3.2$, and $2.3$ day.
Particularly notable is the CSM configuration where $R_0 = 4 \times 10^{14} \, {\rm cm}$ (blue). This case shows a marginally optically-thick CSM, whose initial optical depth $\tau_0 \simeq c/v_{\rm sh}$ is at the critical value below which radiation would not be trapped.
The optical light-curve in this case is especially sensitive to radiative losses during the initial planar expansion phase. This is treated in detail in our current model, but is absent from previous work.
As shown by Fig.~\ref{fig:Piro_Comparison}, this can dramatically impact the observed light-curve within the optical band.

Fig.~\ref{fig:Piro_Comparison} also shows that even in cases where the models qualitatively agree in their optical light-curves, the early bolometric luminosity differs substantially between the two. This would be particularly important once short-cadence UV observations of such events become prevalent.
The \cite{Piro15} model, just like the \cite{Arnett80} SN model, implies a flat light-curve at times $t \ll \ta$. In contrast our current model results in a more complex temporal behavior at $t \lesssim \ta$ (e.g. Fig.~\ref{fig:LC_shape}), a consequence of: (1) that we solve the full spatially-dependent radiative diffusion problem, (2) our treatment of the planar-phase, and (3) the fact that we impose the physical initial condition $L(t=0) = L_{\rm sh}$, i.e. that the initial luminosity equal the shock (breakout) luminosity.
Future wide-field high-cadence UV surveys such as ULTRASAT \citep{Sagiv+14} will be able to directly probe this regime, in which our current model's predictions diverge more noticeably from several previous works.

\section{Discussion}
\label{sec:discussion}

In this work we have derived an analytic solution for the bolometric light-curve emitted by a dense shocked CSM shell, starting from shock-breakout and through subsequent expansion and cooling-envelope emission.
For the first time, we account for geometrical aspects of this problem, which differ from standard SNe models at early times when the shocked CSM more resembles a thin shell than a sphere.
This planar phase has been discussed also in previous work (e.g. \citealt{Nakar&Sari10,Piro+10,Sapir+11,Katz+12,Faran&Sari19}), however only in the regime where $t \ll t_a$ (before the diffusion front crosses the bulk of the material). Here we treat the general case and provide a full analytic solution applicable at all times following shock-breakout. This novel aspect is particularly important in describing the light-curves of ``marginal'' CSM with optical depths $\gtrsim c/v_{\rm sh}$.
We solve the temporal and spatially dependent diffusion equation (eq.~\ref{eq:second_law}) by separation of variables and obtain a solution to the CSM internal energy density as a series expansion in eigenfunctions (eq.~\ref{eq:e}). Imposing appropriate boundary and initial conditions, we obtain a closed-form solution to the radiated luminosity as a function of time (eq.~\ref{eq:L}; see also~\ref{eq:Appendix_L_full}). This result is relevant to the emerging class of fast optical transients (e.g. FBOTs) that are thought to be powered by circumstellar interaction of this sort,
and potentially also to SLSNe-II and to the first peak of double-peaked Type IIb SNe.
We discussed properties of our solutions in \S\ref{sec:properties}, and compared our results to previous models (\S\ref{sec:observations}).

Our work can be seen as an extension to SNe models, where only spherical geometry has been considered \citep[e.g.][]{Arnett79,Arnett80,Pinto&Eastman00}.
Here we have argued that a dense shocked CSM is initially better described by planar degrees of freedom, and that these more appropriately determine the relevant eigenvalues for such CSM.
Formally, this amounts to our neglect of the spherical curvature term, proportional to $S(t) \ll 1$ at early times (eq.~\ref{eq:second_law3}). This is required for eq.~(\ref{eq:second_law3}) to be separable. Although this assumption breaks down deep in the spherical regime ($t \gg \tdyn$), our overarching results remain physically valid and it is only the precise eigenvalues that may not be fully correct within this regime. This can be seen by observing that the one-zone model described in Appendix~\ref{sec:one-zone} results in the same temporal dependence as our detailed solution (eq.~\ref{eq:L}), with only slightly different prefactors (an effectively different eigenvalue; see \S\ref{sec:properties}). The one-zone model does not make any additional assumptions (in particular, there are no terms that are neglected), and therefore lends credence to the physical nature of our solution even within the spherical regime.
Further strengthening this point, we note that in the limit $t, \ta \gg \tdyn$, the light-curve of individual modes reduce to the familiar temporal dependence of the spherical solution (eq.~\ref{eq:L_spherical}; \citealt{Arnett80}) with only a negligible logarithmic correction (eq.~\ref{eq:pwrlaw_spherical_drop}).
The points above also become increasingly less important for CSM shells whose optical depth is only slightly above $\gtrsim c/v_{\rm sh}$, and such CSM may be particularly relevant for observed fast optical transients \citep[e.g.][]{Rest+18}.

Our model provides novel predictions in the regime where the optical depth is $\tau_0 \sim c/v_{\rm sh}$ (or equivalently $\ta \sim \tdyn$; eq.~\ref{eq:tautilde}): it implies a light-curve that can decline more rapidly than previous estimates (Fig.~\ref{fig:Piro_Comparison}). This is due in part to radiative losses that occur already during the planar phase (see discussion surrounding eq.~\ref{eq:Delta_pn}).
Although the light-curve of {\it individual modes} within our eigenfunction expansion revert (qualitatively) to the previously known solutions deep within the spherical regime $\tdyn \ll t$ \citep{Arnett80}, the {\it total} light-curve (eq.~\ref{eq:L}; i.e. the combined ensemble of modes) differs dramatically from these earlier models, and we find a declining luminosity (as roughly $L \propto t^{-1}$; eq.~\ref{eq:dlnLdlnt}) in place of a ``flat'' light-curve at times $\tdyn \ll t \ll \ta$.
This is a consequence of the initial conditions we impose, and that are---for the first time---motivated by the expected shocked CSM configuration at early times. In particular, these initial conditions ensure that our model accurately reproduces the initial breakout flash that is expected in this scenario.

Although we have discussed ``top-hat'' CSM shells throughout the paper, our results are not sensitive to this assumption. We only require that the initial (pre-shock) CSM density profile have an $\sim$abrupt outer edge near $r \sim R_0$, where most of the mass is concentrated and through which the optical depth is $>c/v_0$.
In particular, the model can be applied to the case of a truncated wind profile ($\rho = D r^{-2}$ for $r<R_0$) if we identify $M \approx 4\pi D R_0$ and $\Delta R_{\rm csm} \approx R_0$.

Our analytic solution assumes homologously expanding shocked CSM starting from the time of shock breakout ($t=0$). In practice, homologous expansion is only established on a timescale $\sim 3 \tnot / \sqrt{2} \beta$ over which the rarefaction wave crosses the shocked material. This may introduce deviations in the resulting light-curves over this initial acceleration timescale. We expect such deviations to be modest because: (i) the maximal difference in velocities before and after acceleration is only a modest factor of $\beta^{-1} \sim \mathcal{O}(1)$, and (ii) the true light-curve must converge to our model predictions at both very early times (at $t=0$ the luminosity is set by the breakout luminosity) and late times (at $t \gtrsim \tnot/\beta$ acceleration has negligible and our approximations hold).
Our current model also assumes uniform mass and energy density distributions of the homologously expanding shocked-CSM. The latter are reasonable assumptions for the post-shock CSM at $t=0$, however the (neglected) acceleration phase discussed above may change these distributions. The model can be extended to accommodate arbitrary alternative initial conditions (density and energy-density distributions) though, in general, this may not permit a closed-form analytic solution.

In the present work we have focused on properties of the bolometric luminosity, however we also briefly discuss expectations for the optical-band light-curves in \S\ref{sec:observations}. Treating the emission as blackbody with temperature $T_{\rm eff}$, we show representative optical light-curves in Fig.~\ref{fig:optical}. These indicate that the peak (rise) time is typically set by the temperature evolution, and occurs only once the temperature sweeps through optical band ($T_{\rm eff} \sim h\nu_{\rm opt}/3k_b \sim 10^4\,{\rm K}$). This differs from $^{56}$Ni-powered core-collapse or thermonuclear SNe, whose optical peak is typically set by the photon diffusion timescale \citep{Arnett82,Pinto&Eastman00}. This situation also implies a chromatic dependence of the optical peak, with bluer bands peaking at earlier epochs.
Future planned wide-field high-cadence UV missions (e.g. ULTRASAT; \citealt{Sagiv+14}) would allow observations that more directly probe this scenario.

Actual band-limited light-curves may differ from our simple estimates if the photospheric temperature departs from $T_{\rm eff}$. In particular, at early times it is possible that radiation is out of equilibrium, leading to more complicated spectral dependence \citep[e.g.][]{Nakar&Sari10}. Additionally, H-recombination may affect the light-curve, and this is not currently treated within our present model. For fiducial CSM parameters we typically find that the time of recombination $t_i$ occurs at $t_i > \ta$, in which case recombination should not effect the overall bolometric luminosity \citep{Popov93}. We leave more detailed investigation of these aspects to future work.

One important caveat that may limit applicability of our model, in its present form, to observed transients is the fact that we only treat emission from the shocked CSM. This amounts to neglecting the reverse shock that passes through the material that initially collides with the CSM (the ejecta). Emission from the shocked ejecta may also contribute to the observed light-curve, and this component will be treated separately in upcoming work. In this context, our current results can be viewed as a full solution to the CSM component of the emission. This will accurately describe the combined (CSM+ejecta) light-curve in the limit where 
the ejecta mass is large and $\rho_{\rm ej} \gg \rho_{\rm csm}$. The reverse shock kinetic power is then smaller by a factor $\sim (\rho_{\rm ej}/\rho_{\rm csm})^{-1/2} \ll 1$ than the forward shock power $L_{\rm sh}$ (eq.~\ref{eq:L_0}).
A related caveat is the fact that we assume the CSM-shocked light-curve is powered entirely by the initial energy deposition of the shock. This neglects the possibility of late-time reheating of this material e.g. by radioactive $^{56}$Ni. While there is no reason to expect $^{56}$Ni within the CSM itself, radioactivity within the ejecta could power a secondary peak at late times \citep[e.g.][]{Nakar&Piro14}.

Finally, we note again that many of the novel features derived in our present work are enhanced for ``marginal'' CSM, with optical depth not dramatically larger than $c/v_{\rm sh}$. 
In this regime there are several additional interesting questions that arise and which motivate future work. Most importantly, the width of a radiation mediated shock in such marginal CSM becomes comparable to the total CSM shell width \citep{Weaver76}, which renders steady-state shock solutions questionable. In fact, it is not totally obvious that a radiation-mediated shock is even able to develop for marginal $\tau_0$ because it must take a finite amount of time to transition from a collisionless shock (that must necessarily form at the time of first contact between ejecta and CSM) to one dominated by radiation-pressure forces.


\acknowledgements
I thank Eliot Quataert, Anna Ho, and the anonymous referee for discussions and comments that helped improve this manuscript.
BM is supported by NASA through the NASA Hubble Fellowship grant \#HST-HF2-51412.001-A awarded by the Space Telescope Science Institute, which is operated by the Association of Universities for Research in Astronomy, Inc., for NASA, under contract NAS5-26555.

\appendix

\section{One-zone Model}
\label{sec:one-zone}

The solution obtained in the main text accounts for the spatial distribution of internal energy within the shocked CSM, leading to the partial differential equation~(\ref{eq:second_law}). 
Here we show that the main qualitative results of our work can also be derived within a simplified one-zone model. Below, we recap this model for completeness, and compare its results to the more detailed analysis presented in this paper.

The {\it volume integrated} equivalent of eq.~(\ref{eq:second_law}) can be written as
\begin{equation}
\label{eq:Appendix_E_equation}
\frac{dE}{dt} = -\frac{1}{3} \frac{E}{V} \frac{dV}{dt} - \frac{E}{t_{\rm diff}} .
\end{equation}
This describes the temporal energy evolution of homologously expanding material whose {\it total} internal energy is $E(t)$. The terms on the RHS account for adiabatic and radiative losses, respectively, so that $L = E/t_{\rm diff}$ is the radiated luminosity.

A one-zone equation of this sort is commonly used throughout the literature \citep[e.g][]{Arnett79,Kasen&Bildsten10,Piro15}, traditionally in the spherical limit where $\dot{V}/V = 3/t$. This results in the well-known scalings, first derived by \cite{Arnett79}, that $L \sim \left( E_0 \tdyn / \ta^2 \right) e^{-t^2 /2 \ta^2}$.
Here, we extend these models by accounting for the full time-evolution of the $V(t)$, as relevant to the shocked-CSM scenario. This is explicitly given in eq.~(\ref{eq:V})

The diffusion time throughout {\it the bulk} of the CSM can be expressed as
\begin{align}
\label{eq:Appendix_t_diff}
t_{\rm diff} 
&= \frac{\kappa M}{c V(t)} \left[ R(t) - R_{\rm in}(t) \right]^2
\\ \nonumber
&= \frac{\ta^2 \left[\tnot + (1-\beta)t \right]^2}{\left( \tdyn + \tnot + t \right)^3 - \left( \tdyn + \beta t \right)^3}
.
\end{align}
Inserting eqs.~(\ref{eq:V},\ref{eq:Appendix_t_diff}) into eq.~(\ref{eq:Appendix_E_equation}), we obtain an ODE for $E(t)$ that can be integrated by separation of variables.
The full time-dependent solution, expressed in terms of the luminosity, $L = E(t)/t_{\rm diff}(t)$, is
\begin{align}
\label{eq:Appendix_OneZone}
L(t) 
= E_0 
&\frac{ \left[ \left( \tdyn + \tnot + t \right)^3 - \left( \tdyn + \beta t \right)^3 \right]^{\frac{2}{3}} }{ \ta^2 \left( \tnot + (1-\beta)t \right)^2 }
\\ \nonumber
&\times
\left[ \left( \tdyn + \tnot \right)^3 - \tdyn^3 \right]^{\frac{1}{3}}
\\ \nonumber
&\times
\left[ 1 + (1-\beta)\frac{t}{\tnot} \right]^{-3 \left(\frac{\tdyn}{\ta}\right)^2 \frac{\left( 1-\beta-\beta \tnot/\tdyn \right)^2}{\left(1-\beta\right)^3}}
\\ \nonumber 
&\times
e^{-\frac{t \left[ (1-\beta^3)t + \left( 2 - 4\beta(\beta+1)\right)\tnot + 6(1-\beta^2)\tdyn \right]}{2 (1-\beta)^2 \ta^2}} 
.
\end{align}
Equation~(\ref{eq:Appendix_OneZone}) results in qualitative similar behaviour as that of the more detailed model presented in the main text. One of the main quantitative differences relates to the numerical coefficients in the exponential terms.
In the spatially-dependent problem, these coefficients are proportional to the eigenvalues of different spatial modes, $\alpha_n$ (eq.~\ref{eq:alpha_n}). The one-zone model only appropriately describes the light-curve decay rate if $\alpha_n = 3$. Given the derived eigenvalues, this is not a terrible approximation for the fundamental mode, for which $\alpha_1 \simeq 2.47$. However, this does not adequately described higher-order modes.
This tension between our one-zone model and the full spatially-dependent model is similar to that discussed by \cite{Khatami&Kasen19} in the context of the standard Arnett model \citep{Arnett80,Arnett82}.
Furthermore, the one-zone model is not properly normalized at early times where we expect the luminosity to plateau to the breakout (shock) luminosity, as imposed by the initial conditions in \S\ref{sec:diffusion} (see Fig.~\ref{fig:modes}).
We discuss these points further in \S\ref{sec:modes}.

\section{Complete Light-Curve Expression}
\label{sec:explicit_LC}

For convenience, we below provide the analytic light-curve solution (eq.~\ref{eq:L}) written explicitly in terms of the fundamental variables: $\tnot$, $\tdyn$, $\ta$, $\beta$, and $E_0$ (see Tab.~\ref{tab:variables} for definitions of these variables).
It is
\begin{widetext}
\begin{align}
\label{eq:Appendix_L_full}
    L(t) 
    &= \frac{7^2}{6^3}\beta \frac{E_0}{\tdyn^2} 
    \frac{\left( \tdyn + \tnot + t \right)^2}{ \tnot + (1-\beta)t }
    \left[ \frac{\left( \tdyn + \tnot + t \right)^3 - \left( \tdyn + \beta t \right)^3}{\left( \tdyn + \tnot \right)^3 - \tdyn^3} \right]^{-1/3}
    \sum_{n=1}^{\infty} 
    \frac{4}{(2n-1)\pi}
    \sin \left[ 
    \frac{2n-1}{2}\pi
    \frac{7^2}{6^3} \beta \left(\frac{\tdyn}{\ta}\right)^2 
    \right]
    \\ \nonumber
    &\times
    \left[ 1 + \left(1-\beta\right)\frac{t}{\tnot}\right]^{-\left(\frac{2n-1}{2}\pi\right)^2 \left(\frac{\tdyn}{\ta}\right)^2 \frac{\left(1-\beta-\beta \tnot/\tdyn \right)^2}{1-\beta}}
    \exp\left\{ -\left(\frac{2n-1}{2}\pi\right)^2 \frac{t \left[ (1-\beta^3)t + (2-4\beta(\beta+1))\tnot + 6(1-\beta^2)\tdyn \right]}{6(1-\beta)^2 \ta^2} \right\}
\end{align}
\end{widetext}


\bibliography{bib}{}
\bibliographystyle{aasjournal}



\end{document}